\documentclass[runningheads]{llncs}

\usepackage{graphicx}
\usepackage[numbers,sort&compress]{natbib}
\usepackage{amsmath}
\usepackage{adjustbox}
\usepackage{multirow}
\usepackage{amssymb}
\usepackage{mathtools, nccmath}

\newcommand{\scriptD}{\mathcal{D}}

\newcommand{\scriptL}{\mathcal{L}}
\newcommand{\scriptA}{\mathcal{A}}

\DeclarePairedDelimiter{\nint}\lfloor\rceil
\begin{document}
\title{Bag of Tricks for Developing Diabetic Retinopathy Analysis Framework to Overcome Data Scarcity}
\titlerunning{Bag of Tricks for Diabetic Retinopathy Analysis}
%
\author{Gitaek Kwon \and Eunjin Kim \and Sunho Kim \and Seongwon Bak \and Minsung Kim \and Jaeyoung Kim\thanks{Correspondence to Jaeyoung Kim}}

%
\authorrunning{Gitaek Kwon et al.}
%
\institute{VUNO Inc., Seoul, South Korea \\
\email{\{gitaek.kwon,eunjin.kim,ksunho0660,seongwon.bak,minsung.kim, jaeyoung.kim\}@vuno.co}}
\maketitle              
\begin{abstract}

Recently, diabetic retinopathy (DR) screening utilizing ultra-wide optical coherence tomography angiography (UW-OCTA) has been used in clinical practices to detect signs of early DR. 
However, developing a deep learning-based DR analysis system using UW-OCTA images is not trivial due to the difficulty of data collection and the absence of public datasets.
By realistic constraints, a model trained on small datasets may obtain sub-par performance.
Therefore, to help ophthalmologists be less confused about models' incorrect decisions, the models should be robust even in data scarcity settings.
To address the above practical challenging, we present a comprehensive empirical study for DR analysis tasks, including lesion segmentation, image quality assessment, and DR grading. 
For each task, we introduce a robust training scheme by leveraging ensemble learning, data augmentation, and semi-supervised learning. 
Furthermore, we propose reliable pseudo labeling that excludes uncertain pseudo-labels based on the model's confidence scores to reduce the negative effect of noisy pseudo-labels.
By exploiting the proposed approaches, we achieved 1st place in the Diabetic Retinopathy Analysis Challenge.


\end{abstract}

\section{Introduction}

Diabetic retinopathy (DR) is an eye disease that can result in vision loss and blindness in people with diabetes, but early DR might cause no symptoms or only mild vision problems \cite{gregori2021diabetic}.
Therefore, early detection and management of DR play a crucial role in improving the clinical outcome of eye condition. 
Color fundus photography, fluorescein angiography (FA), and optical coherence tomography angiography (OCTA) have been used in diabetic eye screening to acquire valuable information for DR diagnosis and treatment planning.
Recently, in the screening, ultra-wide OCTA (UW-OCTA) images have been widely used leveraging their advantages such as more detailed visualization of vessel structures, and ability to capture a much wider view of the retinal compared to previous standard approaches \cite{zhang2018ultra}.

With the advancements of deep learning (DL), applying DL-based methods for medical image analysis has become an active research area in the ophthalmology fields \cite{li2021applications, sarki2020automatic, son2020development, son2019towards}.
Notably, the availability to large amounts of annotated fundus photography has been one of the key elements driving the quick growth and success of developing automated DR analysis tools.
\citet{sun2021lesion} develop the automatic DR diagnostic models using color fundus images, and \citet{zhou2019collaborative} propose a collaborative learning approach to improve the accuracy of DR grading and lesion segmentation by semi-supervised learning on the color fundus photography.
Although previous studies investigate the effectiveness of applying DL to DR grading and lesion detection tasks based on color fundus images, DR analysis tool leveraging UW-OCTA are still under-consideration.
One of the reasons lies in the fact that annotating high-quality UW-OCTA images is inherently difficult because the annotation of medical images requires manual labeling by experts.
Consequently, when we consider about the practical restrictions, it is one of the most crucial things to develop a robust model even in the lack of data.

To address the above real-world setting, we introduce the bag of tricks for DR analysis tasks using the Diabetic Retinopathy Analysis Challenge (DRAC22) dataset, which consists of three tasks (i.e., lesion segmentation, image quality assessment, and DR grading)~\cite{DRAC22}.
To alleviate the negative effect introduced by the lack of labeled data, we investigate the effectiveness of data augmentations, ensembles of deep neural networks, and semi-supervised learning.
Furthermore, we propose reliable pseudo labeling (RPL) that selects reliable pseudo-labels based on a trained classifier's confidence scores, and then the classifier is re-trained with labeled and trustworthy pseudo-labeled data. 

In our study, we find that Deep Ensembles \cite{lakshminarayanan2017simple}, test-time data augmentation (TTA), and RPL have powerful effects for DR analysis tasks.
Our solutions are combinations of the above techniques and achieved 1st place in all tasks for DRAC22.

\section{Related Work}

In this section, we overview previous studies on the DR analysis (Sec. \ref{sec:related_dr}), and semi-supervised learning algorithms (Sec. \ref{sec:related_ssl}).

\subsection{Diabetic Retinopathy Analysis}\label{sec:related_dr}


Automatic DR assessment methods based on neural networks have been developed to assist ophthalmologists \cite{qummar2019deep, ruamviboonsuk2019deep, zhang2019automated}. 
\citet{gulshan2016development} develop the convolution neural network (CNN) for detecting DR, and the proposed method shows the competitive result with ophthalmologists in detection performance. They demonstrates the feasibility of the DL-based computer-aided diagnosis system for fundus photography.  
\citet{dai2021deep} suggest a unified framework called DeepDR in order to improve the interpretability of CNNs. 
DeepDR provides comprehensive predictions, including DR grade, location of DR-related lesions, and an image quality assessment of color fundus photography.

On the other hand, a series of approaches based on the FA \cite{pan2020multi, gao2022end}, OCT \cite{ghazal2020accurate, heisler2020ensemble}, and OCTA \cite{zang2020dcardnet, ryu2021deep} have been studied to detect DR.
\citet{pan2020multi} propose the CNN-based model, which classifies DR findings (i.e., non-perfusion regions, microaneurysms, and laser scars) with FA. 
\citet{heisler2020ensemble} suggest an ensemble network for DR classification. Each ensemble member is trained with OCT and OCTA, respectively. For a testing time, they use aggregated predictions of the ensemble model to provide robust and calibrated predictions. 

Although the previous methods have shown remarkable results in promoting the accuracy of DR grading, a comprehensive empirical study of applying UW-OCTA to DL has yet to be conducted.

\subsection{Semi-supervised Learning}\label{sec:related_ssl}

In the medical imaging domain, collecting labeled data is challenging due to expensive costs and time-consuming. Instead, it is much easier to obtain unlabeled data.
Thanks to the recent success of semi-supervised learning (SSL), various SSL algorithms \cite{berthelot2019remixmatch, pham2021meta} show impressive performance on various tasks such as semantic segmentation, object detection, and image recognition.

Pseudo-labeling (PL) \cite{lee2013pseudo} is a simple and effective method in SSL approaches, in which pseudo labels are generated based on the pretrained-network's predictions, and then the network is re-trained both labeled and pseudo-labeled data simultaneously.
Following the pioneering approach of pseudo-labeling, \citet{sohn2020fixmatch} propose FixMatch, which produces pseudo labels using both consistency regularization and pseudo-labeling. FixMatch only retains a pseudo label if the network produces a high probability for a weak-augmented image in order to reduce an error of the prediction caused by the distortions of a given image.
\citet{xie2020self} suggests an iterative training scheme for SSL, called noisy student training.  
In their process, they first train a model on labeled data and use it as a teacher network to generate pseudo labels for unlabeled data. They then train an equal-or-larger model as a student network on the combination of labeled and pseudo-labeled samples and iterate the above process by assigning the student as the teacher.

\section{Method}

In this section, we first introduce a simple and effective technique, reliable pseudo labeling (RPL), for improving classification performance in the data scarcity setting. Then, we describe our solutions for each task in detail. 

\noindent \textbf{Scope of the paper.}
We consider a standard multi-class classification problem which comprise a common setting in computer vision tasks.
In this paper, our goal is for a model to classify as $\hat{y}_i = \{1, ..., C\} \in \mathbb{N}$ (i.e., discrete random variable) for a given $x_i$.

\subsection{Reliable Pseudo Labeling}

\begin{figure}[t]
\includegraphics[width=\textwidth]{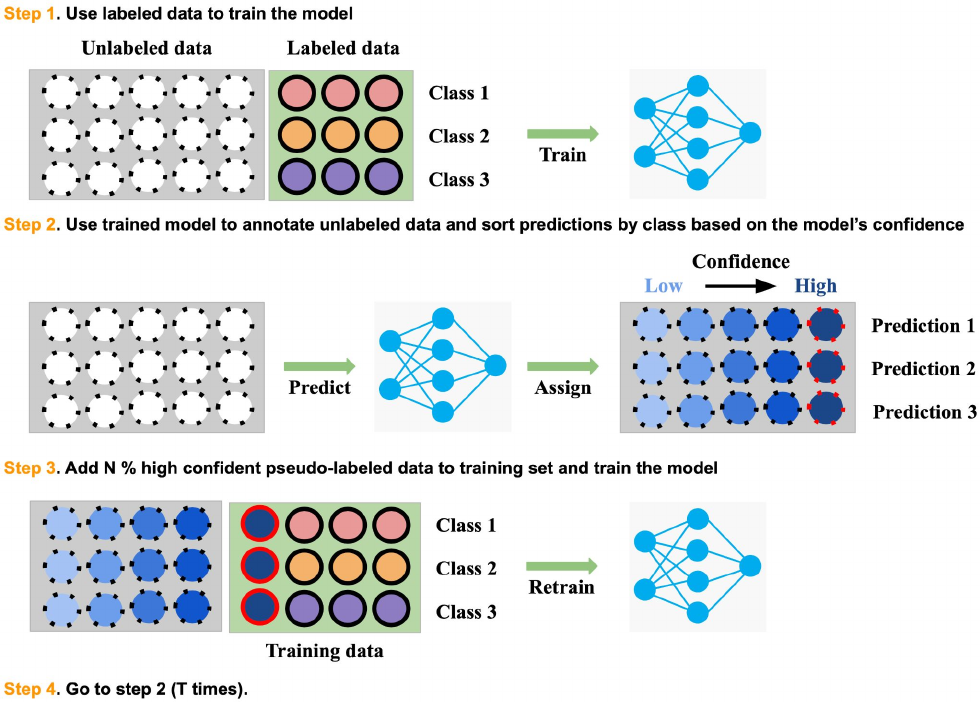}
\caption{Overall procedure of RPL.} \label{fig:rpl}
\end{figure}

\noindent \textbf{Pitfall of pseudo-labels.} 
While PL shows powerful performance in the data scarcity setting, networks can produce an incorrect prediction on unseen data.
If the model is trained using incorrect pseudo-labels, errors accumulate and confirmation bias can appear since modern over-parameterized neural networks easily overfit to noisy samples~\cite{BIAS_PL}.
Hence, it would be reckless to consider all pseudo-labels generated by a network trained with a small amount of data as correct predictions.
To address the vulnerability of PL, we propose RPL, and overall procedure is described in Fig. \ref{fig:rpl}.





\noindent \textbf{Notation.}
Let $[m] := \{1, ..., m\}$, $\sigma(\cdot)$ is the softmax function, and $\nint{\cdot}$ is the nearest integer function.
We denote a given dataset by $\scriptD = \scriptD^l \cup \scriptD^u$, where $\scriptD^l = \{(x_i^l, y_i^l): i \in [n] \}$ is the labeled set and $\scriptD^u = \{x_i^u : i \in [m]\}$ is the unlabeled set.
For multi-class classification tasks, a softmax classifier $f_{\text{cls}}$ maps an input $x_i \in \mathbb{R}^{W \times H}$ into a predictive distribution $\hat{p}(y|\sigma(z_i))$, where $z_i$ is a vector of logits $f_{\text{cls}}(x_i)$ and $y \in [C]$ is a discrete class label. 
When the classification task is formulated by regression problem, a class prediction of a regressor $f_{\text{reg}}$ can be calculated by $\nint{f_{\text{reg}}(x_i)}$ when a label is defined as $y \in  \mathbb{N}$.

\noindent \textbf{Procedure of RPL.}
We first train $f$ using an arbitrary loss function using $\scriptD_{\text{train}}^l$. 
After training, we collect pseudo-labeled set per predicted class $\hat{y}$ defined as $\hat{\scriptD}^u = \{\hat{\scriptD}_k^u\}_{k=1}^C$, where $\hat{\scriptD}_k^u = \{(x^u_i, \hat{y}^u_i)| \hat{y}^u_i = k, i \in [m]\}$.
Then we sort each $\hat{\scriptD}_k$ by the model's confidence score corresponding the predicted class.
For the softmax classifier, the confidence score can be $ \max_c \hat{p}(y=c| \sigma(z_i))$.
When the classification task is formulated as regression, the confidence score can be calculated with $|\nint{f_{\text{reg}}(x_i)}-f_{\text{reg}}(x_i)|$~\footnote{In this paper, we only consider the classification task with a \textit{discrete} label space.}.



To exclude uncertain data, we select $N^t_c$ (\%) confident pseudo labeled samples in each $\hat{\scriptD}_c^u$, and then re-train the model which learns both labeled and trustworthy pseudo-labeled data.
Finally, we repeat the process $T$ times. For each process, $N^t_c$ is increased by $N^t_c = 100 \frac{t \cdot |\scriptD_c^u|}{T}$, where $t = \{1, ..., T\}$ is an indicator of the process.

\subsection{Overview of Solutions}
For readers’ convenience, we provide a brief description of our solutions in Tab. \ref{tab:summary_solutions}.
In the rest of this section, we introduce our solutions in detail.

\begin{table}[]
\caption{Summary of methods used in experiments. Unless otherwise specified, we use reported methods and their parameters for all experiments. MPA \cite{liu2016multi}: Multi-scale patch aggregation.}
\begin{adjustbox}{width=\textwidth,center}
\begin{tabular}{c|c|c|c}
\hline
\textbf{} & \textbf{Task 1} & \textbf{Task 2} & \textbf{Task 3} \\ \hline
\textbf{Preprocessing} & \multicolumn{3}{c}{Dividing all pixels by 255} \\ \hline
\textbf{Input resolution} & \multicolumn{3}{c}{1024 $\times$ 1024} \\ \hline
\textbf{Post-processing} & \begin{tabular}[c]{@{}c@{}} IRMA: false positive removal \\ NP: Dilation (kernel=5) \\ NV: false positive removal \end{tabular} & \begin{tabular}[c]{@{}c@{}}Class-specific \\ thresholding\end{tabular} & \begin{tabular}[c]{@{}c@{}}Post-editing using \\ the segmentation model\end{tabular} \\ \hline
\textbf{\begin{tabular}[c]{@{}c@{}}Test-time \\ augmentation\end{tabular}}  & \begin{tabular}[c]{@{}c@{}} IRMA: rotate=$\{90,180,270,360\}.$ \\ NP: rotate=$\{90,180,270,360\}.$ \\ NV: rotate=$\{90,180,270,360\}, $\\ $MPA=\{1.0, 1.1, 1.2, 1.3, 1.4\}$\end{tabular} & \begin{tabular}[c]{@{}c@{}}Flip=$\{\verb|horizontal|, $\\ $\verb|vertical|\}$\end{tabular} & \begin{tabular}[c]{@{}c@{}}Flip=$\{\verb|horizontal|,$ \\ $\verb|vertical|\}$\end{tabular} \\ \hline
\textbf{Architecture} & $U^2\text{-Net}$ & EfficientNet-b2 & EfficientNet-b2 \\ \hline
\textbf{Pretrained weights} & - & ImageNet & ImageNet \\ \hline
\textbf{Loss} & \begin{tabular}[c]{@{}c@{}}Weighted dice loss, \\ focal loss, \\ binary cross entropy loss\end{tabular} & Smooth L1 & Smooth L1 \\ \hline
\textbf{Optimizer} & AdamW \cite{loshchilov2017decoupled} & AdamW & AdamW \\ \hline
\textbf{Learning rate} & 1e-4 (w/o scheduler) & 2e-4 (w/o scheduler) & 2e-4 (w/o scheduler) \\ \hline
\textbf{Augmentation} & see Tab. \ref{tab:task1_aug} & see Tab. \ref{tab:task2_aug} & see Tab. \ref{tab:task3_aug} \\ \hline
\textbf{Weight decay} & 1e-2 & 1e-2 & 1e-2 \\ \hline
\textbf{SSL} & - & RPL (T=5) & RPL (T=5) \\ \hline
\textbf{Dropout ratio} & 0.0 & 0.2 & 0.2 \\ \hline
\textbf{Batch size} & 2 & 8 & 8 \\ \hline
\textbf{Epochs} & 400 & 150 & 150 \\ \hline
\textbf{Train/Dev split} & 1:0 & 0.8:0.2 & 0.8:0.2 \\ \hline
\textbf{Ensemble} & \begin{tabular}[c]{@{}c@{}}NP: 5 models, \\ IRMA and NV: w/o ensemble \end{tabular} & 5 models & 5 models \\ \hline
\end{tabular}
\end{adjustbox}
\label{tab:summary_solutions}
\end{table}

\subsection{Lesion Segmentation (Task 1)} \label{sec:task1}

\begin{figure}[t]
\includegraphics[width=\textwidth]{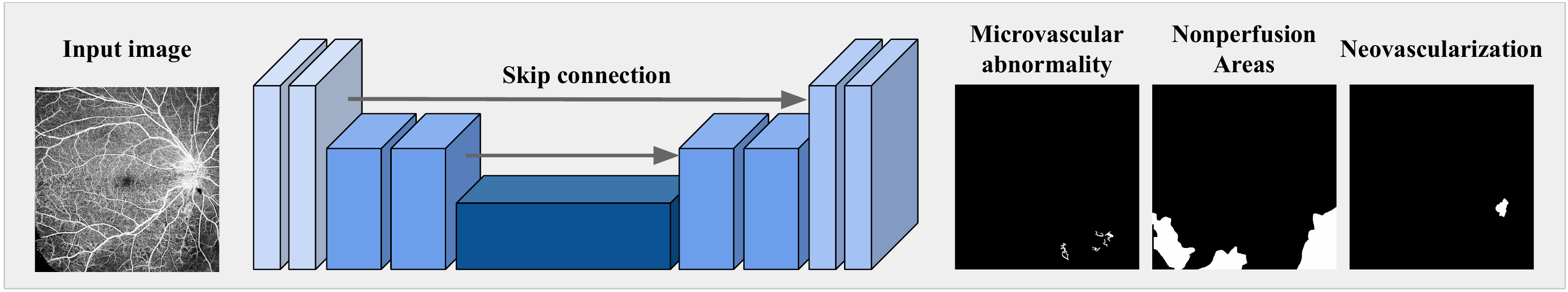}
\caption{For lesion segmentation task, we train the U-shape network $f_{\text{seg}}$. For training, normalized input image $x_i$ is fed into the network, and $f_{\text{seg}}$ are trained with empirical risk minimization.} \label{fig1}
\end{figure}

\noindent \textbf{Motivation.}
The goal of DR-related lesion segmentation task in DRAC22 is detecting the pixel-level lesions, including intraretinal microvascular abnormalities (IRMA), nonperfusion areas (NP), and neovascularization (NV). 
Although one unified segmentation model formulated by multi-label classification can detect the locations of three lesions, the model's detection performance may be sub-optimal since the anatomical characteristics are different among the lesions.
For example, IRMA and NV often appear as small objects, and such imbalanced data distribution may sensitively affects the data-driven deep learning methods \cite{xi2020ia}.
On the other hand, several studies \cite{guo2018mednet} find that a segmentation model confuses NP and signal reduction artifacts. It is natural to think that hard example mining may reduce the false positive for the regions of signal artifacts.
Hence, we design models focusing on imbalanced data setting for small lesions (i.e., IRMA and NV) and hard example mining for NP, respectively.


\noindent \textbf{Training.}
We train two independent $U^2\text{-Net}$ models \cite{qin2020u2}, one is small-lesion segmentation network $f_{\text{seg}}^{\text{small}}$ for IRMA and NV, another is NP segmentation model $f_{\text{seg}}^{\text{NP}}$.
Each model learn to minimize the difference between the predicted lesion maps and the ground-truth masks:
\begin{equation}
    \scriptL_{\text{total}}(y_i, \sigma(f(x_i))) = \scriptL_{\text{Dice}}(y_i, \sigma(f(x_i))) + \alpha \scriptL_{\text{Aux}}(y_i, \sigma(f(x_i))),
\end{equation}
where $\sigma(\cdot)$ is the sigmoid function, $\scriptL_{\text{Dice}}$ is the weighted dice loss, $\scriptL_{\text{Aux}}$ is the auxiliary loss, and $\alpha$ is the hyper-parameter that determines the magnitude of the auxiliary loss.
$\scriptL_{\text{Dice}}$ is calculated by: 
\begin{equation}\label{eq:dice}
    \scriptL_{\text{Dice}}(y, \hat{y}) = 1 - 2 \times \frac{\sum_{c=1}^{C=3} w_c \sum_i y_i^c \cdot \hat{y}_i^c}{\sum_{c=1}^{C=3} w_c \sum_i (y_i^c + \hat{y}_i^c)},
\end{equation}
where $\hat{y} = \sigma(f(x))$ is the prediction given $x$, $C$ is the number of classes, $w_c = \log \frac{1}{\sum_i y_i^c}, i = \{1, ..., W \times H\}$ is the class-wise weight.

We use focal loss \cite{lin2017focal} as $\scriptL_{\text{Aux}}$ for $f_{\text{seg}}^{\text{NP}}$ and modify the original focal loss in order to apply multi-label classification: 
\begin{align}\label{eq:focal}
    \scriptL_{\text{Focal}}(y_i^c, \hat{y}_i^c) =
    \begin{cases}
    - (1 - \hat{y}_i^c) \log \hat{y}_i^c   ,& \text{if } y^c = 1, \\
    - \hat{y}_i^c \log (1 - \hat{y}_i^c)   ,& \text{otherwise}.
    \end{cases}
\end{align}
For $f_{\text{seg}}^{\text{small}}$, we set binary cross-entropy loss as $\scriptL_{\text{Aux}}$ which less penalizes false positive pixels compared to Eq. \ref{eq:focal}. 
In summary, $f_{\text{seg}}^{\text{NP}}$ is trained with strong penalties not only false positive instances but also hard-to-distinguish pixels, whereas $f_{\text{seg}}^{\text{small}}$ is trained focused on positive instances.
Training configurations are summarized in Tab. \ref{tab:summary_solutions}.
We set $\alpha$ as 0.5 for all experiments.

\begin{table}[]
\caption{The list of augmentations for training lesion segmentation models. Each transformation can be easily implemented by albumentation library \cite{info11020125}. The input image has $1024 \times 1024$ resolution.}
\begin{adjustbox}{width=\textwidth,center}
\begin{tabular}{c|c|c}
\hline
\textbf{Operator} & \textbf{Parameters} & \textbf{Probability} \\ \hline
\verb|RandomBrightnessContrast| ($\omega_1$) & \verb|brightness_limit|=0.2, \verb|contrast_limit|=0.2 & \multirow{2}{*}{\textbf{1.0}} \\ \cline{1-2}
\verb|RandomGamma| ($\omega_2$) & \verb|gamma_limit|=(80, 120) &  \\ \hline
\verb|Sharpen| ($\psi_1$) & \verb|alpha|=(0.2, 0.5), \verb|lightness|=(0.5, 1.0) & \multirow{3}{*}{\textbf{1.0}} \\ \cline{1-2}
\verb|Blur| ($\psi_2$) & \verb|blur_limit|=3 &  \\ \cline{1-2}
\verb|Downscale| ($\psi_3$) & \verb|scale_min|=0.7, \verb|scale_max|=0.9 &  \\ \hline
\verb|Flip| ($\phi_1$) & \verb|horizontal|, \verb|vertical| & 0.5 \\ \hline
\verb|ShiftScaleRotate| ($\phi_2$) & \verb|shift_limit|=0.2, \verb|scale_limit|=0.1, \verb|rotate_limit|=90 & 0.5 \\ \hline
\verb|GridDistortion| ($\phi_3$) & \verb|num_steps|=5, \verb|distort_limit|=0.3 & 0.2 \\ \hline
\verb|CoarseDropout| ($\phi_4$) & \begin{tabular}[c]{@{}c@{}} \verb|max_height|=128, \verb|min_height|=32,  \\ \verb|max_width|=128, \verb|min_width|=32, \verb|max_holes|=3 \end{tabular} & 0.2 \\ \hline
\verb|Affine| ($\phi_5$) & \verb|scale|={[}0.8, 1.2{]} & 0.5 \\ \hline
\end{tabular}
\end{adjustbox}
\label{tab:task1_aug}
\end{table}

\noindent \textbf{Augmentation.} For all segmentation models, we use the following data augmentation strategy.
Let $\scriptA = \scriptA_{\text{pixel}} \cup \scriptA^c$ be a set of $n$ augmentations. $\scriptA_{\text{pixel}} = \{\Omega, \Psi\}$ is the subset of $\scriptA$. $\Omega$ and $\Psi$ respectively have child operators, i.e., $\Omega = \{\omega_k\}_{k=1}^m$, and $\Psi = \{\psi_k\}_{k=1}^{m'}$. 
For training the segmentation model, we always apply pixel-wise transformations, and an augmented image is defined as $\bar{x}_i = \Psi^{\ast}(\Omega^{\ast}(x_i))$, where $\Omega^{\ast}$ and $\Psi^{\ast}$ are randomly picked from $\Omega$ and $\Psi$, respectively.   

To generate diverse input representations, we also apply geometric transform $\phi_k$ to $\bar{x}_i$, and $\phi_k$ is randomly sampled from $\scriptA^c = \{\phi_k\}_{k=1}^{n - (m+m')}$.
As a result, segmentation models are trained with $\scriptD_{\text{seg}} = \{(\phi(\bar{x}_i), y_i)\}_{i=1}^{|\scriptD_{\text{seg}}|}$, 
thus, each model never encounter original training samples.
The list of operators and detailed parameters are described in Tab. \ref{tab:task1_aug}.




\noindent \textbf{Ensemble.} To boost the performance, we use ensemble techniques such as TTA and Deep Ensemble~\cite{lakshminarayanan2017simple}.
\begin{itemize}
    \item IRMA: TTA—averaging the predictions of $f_{\text{seg}}^{\text{small}}$ across multiple rotated samples of data—is used.
    \item NP: We use an averaged prediction of five independent models' prediction where each prediction also applied TTA with rotation operators.
    \item NV: TTA with rotation transformation and MPA \cite{liu2016multi} are used.
\end{itemize}

\noindent \textbf{Post-processing.} To reduce the incorrect prediction, the following post-processing methods are used. We denote the prediction masks as $\mathbf{P}_{\text{IRMA}}$, $\mathbf{P}_{\text{NP}}$, and $\mathbf{P}_{\text{NV}}$, respectively.

\begin{itemize}
    \item NP: A prediction mask is applied dilation operation with a kernel size of 5.
    \item IRMA: To reduce the false positive pixels, we replace a positive pixel $p_{ij} \in \mathbf{P}_{\text{IRMA}}$ to negative where $q_{ij} \in \mathbf{P}_{\text{NV}}$ have more confident prediction compared to $p_{ij}$. 
    \item NV: In the same way as above, a positive pixel $q_{ij}$ is replaced with zero when $p_{ij}$ is more confident at the same region. 
\end{itemize}


\subsection{Image Quality Assessment (Task 2) and DR Grading (Task 3)}

\noindent \textbf{Motivation.} The goal of image quality assessment and DR grading task in DRAC22 is to distinguish qualities of UW-OCTA images and grading the severity of DR. 
Image quality consists of three levels: poor quality level (PQL), good quality level (GQL), and Excellent quality level (EQL), and the DR grade consists of three levels: normal, non-proliferatived diabetic retinopathy (NPDR), and proliferatived diabetic (PDR).
We formulate the above tasks as a regression problem rather than a multi-class classification problem in order to consider the correlation among classes.

\noindent \textbf{Training.} For each task, we build the regression model $f_{\text{reg}}$ using the EfficientNet-b2 \cite{tan2019efficientnet} initialized with pretrained weights for ImageNet.
To address the lack of labeled training data, RPL is applied. We use final test set in DRAC22 as unlabeled dataset.
The detailed hyper-parameters and training configurations are reported in Tab. \ref{tab:summary_solutions}.


\begin{table}[]
\caption{The list of augmentations for task 2.}
\begin{adjustbox}{width=\textwidth,center}
\begin{tabular}{c|c|c}
\hline
\textbf{Operator} & \textbf{Parameters} & \textbf{Probability} \\ \hline
\verb|RandomBrightnessContrast| ($\omega_1$) & \verb|brightness_limit|=0.2, \verb|contrast_limit|=0.2 & \multirow{2}{*}{\textbf{1.0}} \\ \cline{1-2}
\verb|RandomGamma| ($\omega_2$) & \verb|gamma_limit|=(80, 120) &  \\ \hline
\verb|Sharpen| ($\psi_1$) & \verb|alpha|=(0.2, 0.5), \verb|lightness|=(0.5, 1.0) & \multirow{3}{*}{\textbf{1.0}} \\ \cline{1-2}
\verb|Blur| ($\psi_2$) & \verb|blur_limit|=3 &  \\ \cline{1-2}
\verb|Downscale| ($\psi_3$) & \verb|scale_min|=0.7, \verb|scale_max|=0.9 &  \\ \hline
\verb|Flip| ($\phi_1$) & \verb|horizontal|, \verb|vertical| & 0.5 \\ \hline
\verb|ShiftScaleRotate| ($\phi_2$) & \verb|shift_limit|=0.2, \verb|scale_limit|=0.1, \verb|rotate_limit|=45 & 0.5 \\ \hline
\end{tabular}
\end{adjustbox}
\label{tab:task2_aug}
\end{table}

\begin{table}[]
\caption{The list of augmentations for task 3.}
\begin{adjustbox}{width=\textwidth,center}
\begin{tabular}{c|c|c}
\hline
\textbf{Operator} & \textbf{Parameters} & \textbf{Probability} \\ \hline
\verb|RandomBrightnessContrast| ($\omega_1$) & \verb|brightness_limit|=0.2, \verb|contrast_limit|=0.2 & \multirow{2}{*}{\textbf{1.0}} \\ \cline{1-2}
\verb|RandomGamma| ($\omega_2$) & \verb|gamma_limit|=(80, 120) &  \\ \hline
\verb|Sharpen| ($\psi_1$) & \verb|alpha|=(0.2, 0.5), \verb|lightness|=(0.5, 1.0) & \multirow{3}{*}{\textbf{1.0}} \\ \cline{1-2}
\verb|Blur| ($\psi_2$) & \verb|blur_limit|=3 &  \\ \cline{1-2}
\verb|Downscale| ($\psi_3$) & \verb|scale_min|=0.7, \verb|scale_max|=0.9 &  \\ \hline
\verb|Flip| ($\phi_1$) & \verb|horizontal|, \verb|vertical| & 0.5 \\ \hline
\verb|ShiftScaleRotate| ($\phi_2$) & \verb|shift_limit|=0.2, \verb|scale_limit|=0.1, \verb|rotate_limit|=45 & 0.5 \\ \hline
\verb|CoarseDropout| ($\phi_3$) & \begin{tabular}[c]{@{}c@{}} \verb|max_height|=5, \verb|min_height|=1,  \\ \verb|max_width|=512, \verb|min_width|=51, \verb|max_holes|=5 \end{tabular} & 0.2 \\ \hline
\end{tabular}
\end{adjustbox}
\label{tab:task3_aug}
\end{table}

\noindent \textbf{Augmentation.} The strategy of augmentation is the same as in Sec. \ref{sec:task1}, but task 2 and task 3 have different combinations of operators, respectively. The detailed components are reported in Tab. \ref{tab:task2_aug}, and Tab. \ref{tab:task3_aug}.

\noindent \textbf{Ensemble.} For testing-time, we use an averaged prediction of five independent models' prediction. Also each prediction of the model is generated by TTA with flip operators.

\begin{figure}[t]
\includegraphics[width=\textwidth]{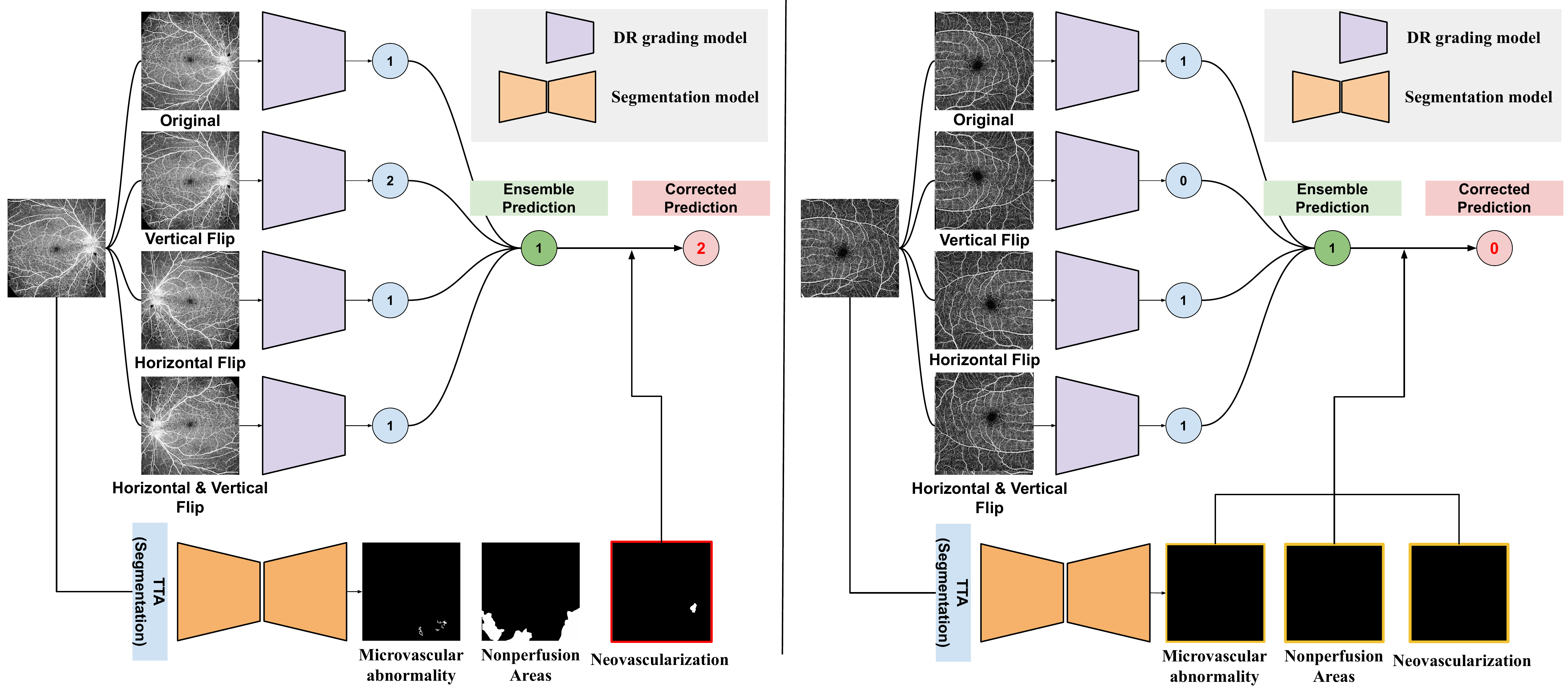}
\caption{Summarization of the post-processing process of task 3. The lesion segmentation model is used to correct incorrect prediction labels.} \label{fig_task3_post}
\end{figure}

\noindent \textbf{Post-processing.} The following post-processing methods are used.
\begin{itemize}
    \item Task 2: We use the following decision rule using class-specific operating thresholds instead of $\hat{y}=\nint{f_{\text{reg}}(x)}$:
    \begin{align}
    \hat{y} = \begin{dcases*}
        0, & if $ f_{reg}(x) < 0.54 $,\\
        1, & if $ 0.54 \leq f_{reg}(x) < 1.5 $,\\
        2, & otherwise. 
        \end{dcases*}
  \end{align}
    \item Task 3: Retrospectively, we find that the DR grading model ignores the NV region (NV is a sure sign of PDR) when the region is small in PDR samples. In this case, the model ultimately misclassifies PDR as NPDR, thus, we replace the failure prediction for NPDR to PDR using the segmentation model's prediction mask of NV (left in Fig. \ref{fig_task3_post}).
    In contrast, if the segmentation model predicts normal for all classes, we correct the the DR grading model's prediction to normal (right in Fig. \ref{fig_task3_post}). 
\end{itemize}

\section{Experiments}

\subsection{Dataset and Metrics}

\setlength{\tabcolsep}{1em}
{\renewcommand{\arraystretch}{1.2}
\begin{table}[]
\caption{Data statistics for DRAC22. \textbf{PQL}: Poor Quality Level, \textbf{GQL}: Good Quality Level, \textbf{EQL}: Excellent Quality Level, \textbf{NPDR}: Non-Proliferatived DR, \textbf{PDR}: Proliferatived DR.}
\begin{adjustbox}{width=8cm,center}
\begin{tabular}{c|cccc|c}
\hline
 & \multicolumn{4}{c|}{\# train} & \# test (unlabeled) \\ \hline
\multirow{2}{*}{Task1} & \multicolumn{1}{c|}{Total} & IRMA & NP & NV & \multirow{2}{*}{65} \\ \cline{2-5}
 & \multicolumn{1}{c|}{109} & 86 & 106 & 35 &  \\ \hline
\multirow{2}{*}{Task2} & \multicolumn{1}{c|}{Total} & PQL & GQL & EQL & \multirow{2}{*}{438} \\ \cline{2-5}
 & \multicolumn{1}{c|}{665} & 50 & 97 & 518 &  \\ \hline
\multirow{2}{*}{Task3} & \multicolumn{1}{c|}{Total} & Normal & NPDR & PDR & \multirow{2}{*}{386} \\ \cline{2-5}
 & \multicolumn{1}{c|}{611} & 329 & 212 & 70 &  \\ \hline
\end{tabular}
\end{adjustbox}
\label{tab:dataset}
\end{table}}

\begin{figure}[t]
\centering
\includegraphics[width=9cm]{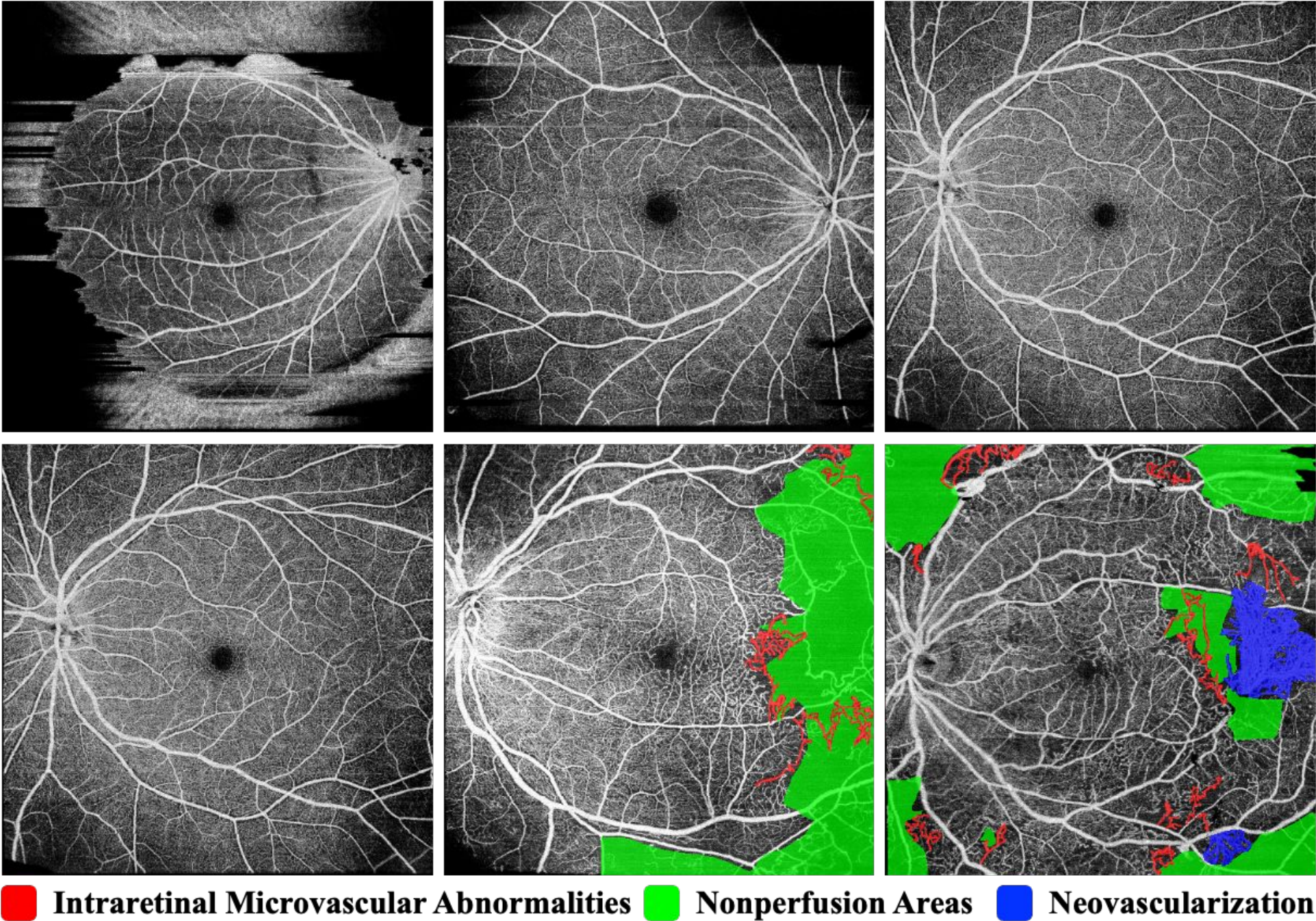}
\caption{Visualized example of DRAC22 dataset. (\textbf{Top}) Each column respectively represents PQL, GQL, and EQL with respect to task 2. (\textbf{Bottom}) In order from left to right, each image represents Normal, NPDR, and PDR, respectively.} \label{fig:example}
\end{figure}

\noindent \textbf{Dataset.} 
In DRAC22, the dataset consists of three tasks, i.e., lesion segmentation, image quality assessment, and DR grading~\footnote{\url{https://drac22.grand-challenge.org/}}.
Data statistics are described in Tab. \ref{tab:dataset}, and we also provide an example of DRAC22 dataset (see Fig. \ref{fig:example}). 
We use 20\% data as a validation set for task 2 and task 3, and we select the best models when validation performance is highest.
Partially, for task 1, we use all training samples so that we select the model which has the highest dice score with respect to the training set.

\noindent \textbf{Metrics.}
For task 1, the averaged dice similarity coefficient (mean-DSC) and the averaged intersection of union (mean-IoU) are measured to evaluate the segmentation models.
For task 2 and task 3, the quadratic weighted kappa (QWK) and Area Under Curve (AUC) are used to evaluate the performance of the proposed methods.



\subsection{Results}
\vspace{-0.3cm}
\setlength{\tabcolsep}{1em}
{\renewcommand{\arraystretch}{1.3}
\begin{table}[]
\caption{The ablation study results for task 1 on testset of DRAC22. Note that ensemble is only used for $f_{\text{seg}}^{\text{NP}}$. Post: Post-processing.}
\begin{adjustbox}{width=12cm,center}
\begin{tabular}{ccc|cc|ccc}
\hline
Ensemble & TTA & Post & mean-DSC & mean-IOU & IRMA DSC & NP DSC & NV DSC \\ \hline
 &  &  & 0.5859 & 0.4311 & 0.4596 & 0.6803 & 0.6179 \\
\checkmark &  &  & 0.5865 & 0.4380 & 0.4596 & 0.6821 & 0.6179 \\
\checkmark & \checkmark &  & 0.5927 & 0.4418 & 0.4607 & 0.6911 & 0.6263 \\ \hline
\checkmark & \checkmark & \checkmark & \textbf{0.6067} & \textbf{0.4590} & \textbf{0.4704} & \textbf{0.6926} & \textbf{0.6571} \\ \hline
\end{tabular}
\end{adjustbox}
\label{tab:ablation_task1}
\end{table}}

\noindent \textbf{Task 1.}
We report the lesion segmentation performance in Tab. \ref{tab:ablation_task1}.
Our best segmentation model achieves the mean-DSC of 0.6067, the mean-IOU of 0.4590, respectively.
Notably, combining ensemble techniques and post-processing indeeed shows the effectiveness for the lesion segmentation task.

\setlength{\tabcolsep}{1em}
{\renewcommand{\arraystretch}{1.3}
\begin{table}[]
\caption{The ablation study results for task 2 on testset of DRAC22.}
\begin{adjustbox}{width=8cm,center}
\begin{tabular}{ccccc|cc}
\hline
Ensemble & PL & RPL & TTA & Post & QWK & AUC \\ \hline
 &  &  &  &  & 0.7321 &  0.7487 \\
\checkmark &  &  &  &  & 0.7485 &  0.7640 \\
\checkmark & \checkmark &  &  &  & 0.7757  & 0.7786 \\
\checkmark &  & \checkmark &  &  & 0.7884  & 0.7942 \\
\checkmark &  & \checkmark & \checkmark &  & 0.7920  & 0.7923 \\ \hline
\checkmark &  & \checkmark & \checkmark & \checkmark & \textbf{0.8090}  & \textbf{0.8238} \\ \hline
\end{tabular}
\end{adjustbox}
\label{tab:ablation_task2}
\end{table}}


\noindent \textbf{Task 2.} 
Tab. \ref{tab:ablation_task2} presents an ablation study to assess each component of the solution by removing parts as appropriate. In Tab. \ref{tab:ablation_task2}, the first row is the performance of baseline, which obtain the QWK of 0.7332 and AUC of 0.8492, respectively.
Other components are incrementally applied, where performance is enhanced consistently with the addition of each step.
Specially, RPL approaches show the noticeable performance improvement with respect to QWK compared to PL.

\setlength{\tabcolsep}{1em}
{\renewcommand{\arraystretch}{1.3}
\begin{table}[]
\caption{The ablation study results for task 3 on testset of DRAC22.}
\begin{adjustbox}{width=8cm,center}
\begin{tabular}{ccccc|cc}
\hline
Ensemble & PL & RPL & TTA & Post & QWK & AUC \\ \hline
 &  &  &  &  & 0.8272 & 0.8749 \\
\checkmark &  &  &  &  & 0.8557 & 0.8856 \\
\checkmark & \checkmark &  &  &  & 0.8343 & 0.8630 \\
\checkmark &  & \checkmark &  &  & 0.8607 & 0.8844 \\
\checkmark &  & \checkmark & \checkmark &  & 0.8684 & 0.8865 \\ \hline
\checkmark &  & \checkmark & \checkmark & \checkmark & \textbf{0.8910} & \textbf{0.9147} \\ \hline
\end{tabular}
\end{adjustbox}
\label{tab:ablation_task3}
\end{table}}


\begin{figure}[t]
\centering
\includegraphics[width=\textwidth]{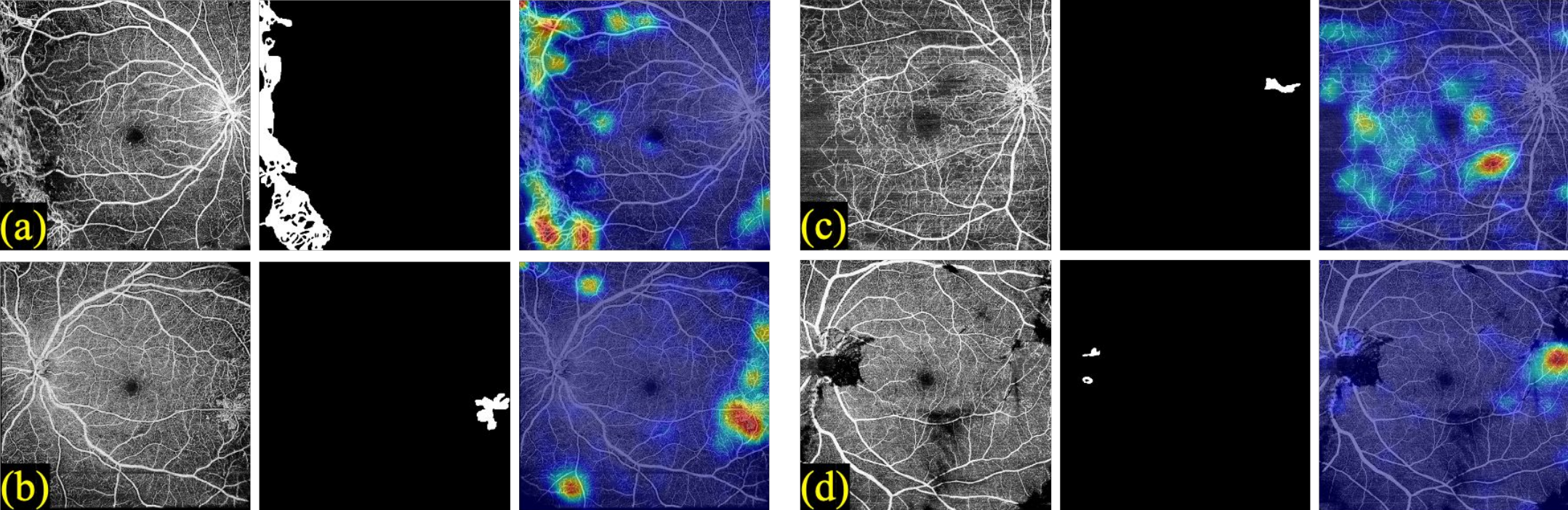}
\caption{Localization result of DR grading model for PDR samples including NV. Each column in the block represents original image, ground truth mask for NV, and activation map, respectively. To obtain the activation map, we use GradCAM \cite{selvaraju2017grad}. (\textbf{a-b}) The prediction result of the model is PDR. (\textbf{c-d}) The prediction result of the model is NPDR (failure cases).   }
\label{fig:c3_gradcam}
\end{figure}

\begin{figure}[t!]
\centering
\includegraphics[width=10cm]{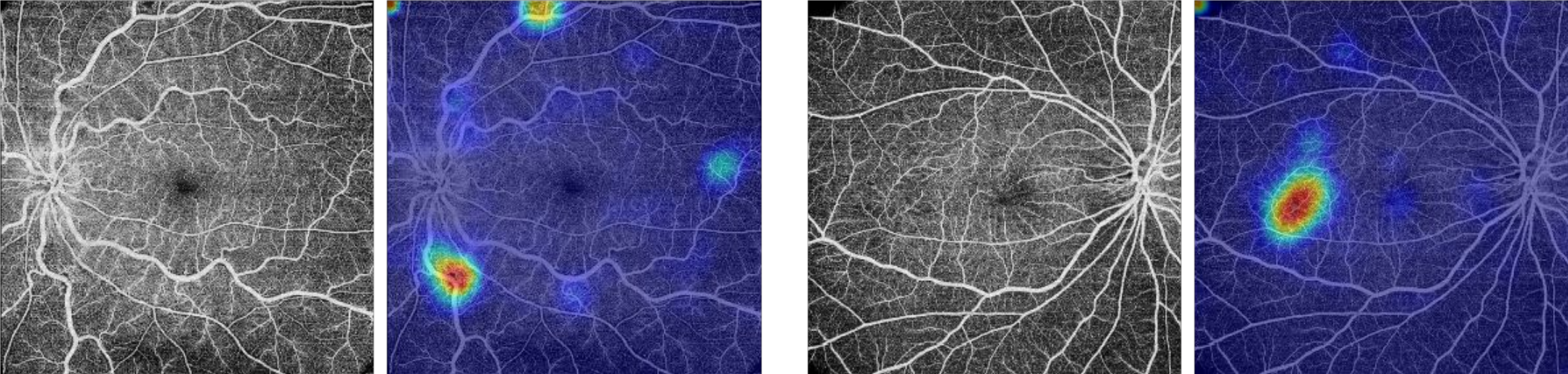}
\caption{Failure cases where the model's prediction is NPDR for normal samples.}
\label{fig:gradcam_zero}
\end{figure}

\noindent \textbf{Task 3.} 
An ablation study is performed with the results shown in Tab. \ref{tab:ablation_task3} to evaluate the performance for multiple techniques. 
The baseline (first row in Tab. \ref{tab:ablation_task3}) obtains sub-par performance compared to other tricks. 
Similar to task 2, model ensemble, RPL, TTA, and the post-editing show advanced performance for QWK and AUC, and applying the post-processing is most effective for DR grading.


To analyze why the post-processing is effective, we analyze what areas the DR grading model is focusing on making predictions.
Fig. \ref{fig:c3_gradcam} shows PDR samples in which the segmentation model detects NV.
In the left block of Fig. \ref{fig:c3_gradcam}, the DR grading model activates the NV and classifies it as PDR when the NV area is large.
On the other hand, when the NV area is small, the DR grading model does not activate the NV and classifies it as NPDR (right in Fig. \ref{fig:c3_gradcam}).
Therefore, it seems useful to correct the results of the DR grading model to PDR when the segmentation model detects NV.
In contrast, as shown in Fig \ref{fig:gradcam_zero}, there are cases where the DR grading model activates the artifact area for normal samples in which the segmentation model detects no lesions.
If the segmentation model does not detect any DR lesions, it seems reasonable to correct the results of the DR grading model to normal.




\section{Conclusion}

In this paper, we present a fully automated DR analysis system using UW-OCTA. We find that various tricks including ensemble learning, RPL, and TTA show advanced performance compared to baselines, and RPL shows significantly improved performance in a data scarcity setting compared to a naive pseudo-labeling. By assembling these tricks, we achieved 1st place in the DRAC22.
We hope our study can revitalize the field of UW-OCTA research and the proposed methods serve as a strong benchmark for DR analysis tasks.
In addition, we expect our approaches to substantially benefit clinical practices by improving the efficiency of diagnosing DR and reducing the workload of DR screening.

\bibliographystyle{splncs04}
\bibliography{reference}





\end{document}